\documentclass[graybox, envcountchap]{svmult}
\usepackage{mathptmx}        % selects Times Roman as basic font
\usepackage{amsmath}
\usepackage{amssymb}
\usepackage{color}
\usepackage{helvet}          % selects Helvetica as sans-serif font
\usepackage{courier}         % selects Courier as typewriter font
\usepackage{dirtree}
\usepackage{makeidx}        % allows index generation
\usepackage{graphicx}        % standard LaTeX graphics tool
\usepackage{subfig}

\usepackage{multicol}        % used for the two-column index
\usepackage[bottom]{footmisc}% places footnotes at page bottom

\usepackage{hyperref}        %for hyperlinks
\hypersetup{colorlinks=true,urlcolor=blue}
\usepackage{calrsfs}
\DeclareMathAlphabet{\pazocal}{OMS}{zplm}{m}{n}
\usepackage[misc]{ifsym}
\usepackage{bm}
\usepackage{bbm}
\usepackage{mathrsfs}
\newcommand{\be}{\begin{eqnarray}}
\newcommand{\ee}{\end{eqnarray}}
\newcommand{\la}{\langle}
\newcommand{\ra}{\rangle}
\newcommand{\bh}{{\rm BH}}

\newcommand{\Tr}{{\rm Tr}}

\newcommand{\del}{\partial}
\def\dg{\dagger}
\begin{document}
\title{Paradox No More: How Stimulated Emission of Radiation Preserves Information Absorbed by Black Holes}
\titlerunning{Stimulated Emission Preserves Information in Black Holes}
%Title is no good: Evaporation has nothing to do with stimulated emission
%Should instead mention the "so-called second law"?  

% Use \titlerunning{Short Title} for an abbreviated version of
% your contribution title if the original one is too long
\author{Christoph Adami}
% Use \authorrunning{Short Title} for an abbreviated version of
% your contribution title if the original one is too long
\institute{C. Adami (\Letter) \at Department of Physics and Astronomy, Michigan State University, East Lansing, MI 48824, \email{adami@msu.edu}
}
%
% Use the package "url.sty" to avoid
% problems with special characters
% used in your e-mail or web address
%
\maketitle

\abstract{Black holes have been implicated in two paradoxes that involve apparently non-unitary dynamics. According to Hawking's theory, information that is absorbed by a black hole is destroyed, and the originally pure state of a black hole is converted to a mixed state upon complete evaporation. Here we address one of the two, namely the apparent loss of (classical) information when it crosses the event horizon. We show that this paradox is due to a mistake in Hawking's original derivation: he ignored the contribution of the {\em stimulated} emission of radiation that according to Einstein's theory of blackbody radiance must accompany the spontaneous emission (the Hawking radiation). Resurrecting the contribution of stimulated emission makes it possible to calculate the (positive) classical information transmission capacity of black holes, which implies that information is fully recoverable from the radiation outside the black hole horizon.} 

\section{Historical Background}
\label{sec:1}
In September of 1973, Stephen Hawking traveled to Moscow to meet the eminent Soviet physicist Yakov Zeldovitch and his student, Alexei Starobinskii. At the time, Hawking had established a classical ``area law" for black holes~\cite{Hawking1971,Hawking1972} that implied that the surface area of black holes can never decrease, but was irritated by Jacob Bekenstein's claim that this area law was the gravitational equivalent of the second law of thermodynamics, by identifying the surface area of the black hole with an entropy~\cite{Bekenstein1972,Bekenstein1973}. Hawking disagreed with Bekenstein's identification~\cite{Hawking1988} because it would imply that black holes have a temperature. Zeldovitch and Starobinskii, however, convinced Hawking that Bekenstein was right. They had been working on a phenomenon called ``superradiance": the amplification of incoming radiation by bulk matter. Zeldovitch had shown that a rotating cylinder would emit such radiation~\cite{Zeldovitch1972}, while Starobinskii applied the same thinking to rotating black holes~\cite{Starobinski1973}.  Zeldovich and Starobinskii suggested to Hawking that black holes would emit radiation not only when impacted by incoming radiation, but {\em also} when stimulated by vacuum fluctuations. After returning from Moscow, Hawking embarked on the calculation that we all now know~\cite{Hawking1975}, showing that black holes indeed emit radiation spontaneously (as if stimulated by the vacuum). 

The phenomenon that Starobinskii described is a classical phenomenon: even water waves scattering off of a rotating flow can create superradiance~\cite{Torresetal2017}. It is a process that leads to the {\em amplification} of incoming waves, and requires incoming waves of {\em negative} effective frequency (or, in the case of superradiance in atoms, a population inversion)\footnote{Radiation can also be amplified when a charged particle moves through a dielectric medium at a speed larger than the speed of light in the medium. This form of superradiance, which obtains its energy from the medium, is known as the \v{C}erenkov effect, see for example~\cite{BekensteinSchiffer1998}.}. There is, however, a {\em quantum} version of superradiance that does not require negative frequencies: this is Einstein's {\em stimulated emission of radiation}~\cite{Einstein1917}. The effective frequency of radiation emanating from a black hole in curved-space quantum field theory is
\be
\omega=\omega_0-m\Omega-\epsilon\Phi\;. \label{omega}
\ee
Here, $\omega_0$ is the frequency of the mode, $m$ is the black hole's orbital angular momentum quantum number, $\epsilon$ the charge, and $\Omega$ and $\Phi$ are the rotational frequency and the electric potential of the black hole, respectively.  For non-rotating uncharged black holes $\omega\geq0$, and such black holes cannot superradiate in the classical form (which requires $\omega<0$).  As a consequence Hawking, apparently unaware of ``quantum superradiance", ignored stimulated emission completely. As we will see, this is the mistake that gave rise to the information paradox.

It turns out that Schwarzschild black holes (that are uncharged and do not rotate) {\em can} emit stimulated radiation: the energy is provided by the black hole just as the energy in Hawking radiation is provided by the black hole. 
Intuitively, just like vacuum fluctuations can create particle--anti-particle pairs out of the vacuum which, if the process takes place close to the horizon, leads to Hawking radiation (spontaneous emission of radiation), an incoming particle can stimulate the creation of a particle--anti-particle pair with quantum numbers aligned with the incoming particle. In modern language, an incoming quantum state will stimulate the emission of a {\em clone}-{\em anti-clone} pair at the horizon~\cite{Simon2000}. This does not violate the quantum no-cloning theorem~\cite{WoottersZurek1982,Dieks1982} because (as argued by Mandel shortly after the publication of the no-cloning theorem~\cite{Mandel1983}) the concommittant spontaneous emission (here, Hawking radiation) will always lead to {\em approximate} cloning, which is not forbidden. 

It was Bekenstein himself who first pointed out Hawking's mistake, in a paper where he treated black holes as ordinary black bodies. Using purely statistical arguments, Bekenstein together with his student Amnon Meisels showed that spontaneous radiation without concommittant stimulated emission is inconsistent, and using Einstein's A- and B-coefficient formalism was able to derive the probability distribution of stimulated radiation simply by insisting on probability conservation~\cite{BekensteinMeisels1977}.  A detailed calculation of the response of Schwarzschild black holes to incoming quanta using Hawking's formulation of semi-classical curved-space quantum field theory~\cite{PanangadenWald1977,AudretschMueller1992,AdamiVersteeg2014} reproduced the Bekenstein-Meisels result precisely. 

But does the stimulated emission effect really save information from destruction in black holes? Not everyone was convinced. Schiffer, for example, argued that spontaneous emission might still ``overpower" the stimulated process, leading to a loss of information overall~\cite{Schiffer1993}. Here I show using a rigorous application of quantum information theory that the information transmission capacity of black holes is strictly positive, which implies that information can be {\em perfectly} reconstructed from the radiation outside of the event horizon of a black hole after information has been absorbed by the black hole. This information is {\em not} encoded in the Hawking radiation, however, but instead can be extracted from the stimulated radiation only, that is, from the almost perfect {\em clones} (copies) of the incoming states~\cite{AdamiVersteeg2015}. 

In the following sections, I discuss the probability distributions of outgoing spontaneous and stimulated emission, outline the calculation of the capacity of quantum black holes to transmit classical information, and discuss its relevance in understanding microscopic time-reversal invariance in black hole dynamics. I close with a brief discussion of the remaining paradox: whether or not the evaporation of a black hole that was formed from a pure state will return to a pure state.

\section{Probability distributions}
The probability distribution of outgoing quanta---both spontaneous and stimulated---can be obtained via a variety of methods. I disuss the simplest of methods first (maximum entropy analysis), and briefly discuss how the same distributions are obtained in curved-space quantum field theory, before calculating the channel capacity. 

\subsection{Spontaneous emission}
The probability distribution $p(n)$ of quanta in any of $n$ outgoing modes emitted by a black hole via spontaneous emission (the distribution of Hawking radiation), can be derived using only Hawking's result that the mean number of outgoing quanta is~\cite{Hawking1975}
\be
\la n\ra=\frac{\Gamma}{e^{x}-1}\;. \label{hawking}
\ee
Here, $\Gamma$ is the absorption coefficient of the black hole (the ``grey-body factor", so that $1-\Gamma$ is the black hole's reflectivity), and $x=\omega/T_\bh$ where $\omega$ is the effective frequency shown in (\ref{omega}) and $T_\bh$  is the black hole temperature.

You can derive the result (for a single massless scalar bosonic mode $n$)~\cite{Bekenstein1975} 
\be
p(n)=(1-e^{-\lambda})e^{-n\lambda}  \label{distrib}
\ee
simply by demanding that the Shannon entropy of the outgoing radiation
\be
S=-\sum_n p(n)\log p(n) \label{shannon}
\ee
is maximal, with the constraints $\sum_n p(n)=1$ and $\sum_n np(n)=\frac {\Gamma}{e^{x}-1}$ implemented via Lagrange multipliers. In Eq.~(\ref{distrib}), the Lagrange multiplier $\lambda$ is related to black hole parameters via
\be
e^{-\lambda}=\frac{\Gamma}{e^x-1+\Gamma}\;.
\ee
Note that distribution (\ref{distrib}) (which was not given by Hawking) was also derived independently using full-fledged curved-space quantum field theory by Wald~\cite{Wald1975}. In the following, we will restrict ourselves to non-rotating and uncharged (Schwarzschild) black holes.  Let us also introduce convenient parameters\footnote{These parameters originate from the curved-space quantum field-theoretic treatment we will discuss later, but serve here to create particularly simple expressions.}  $\alpha$, $\beta$, and $\gamma$ with the definitions
\be
e^{x}&=&\frac{\alpha^2}{\beta^2}\;, \\
\gamma^2&=&1-\Gamma\;,
\ee
with the normalization condition
\be
 \alpha^2-\beta^2+\gamma^2=1\;. \label{norm}
\ee
In terms of these parameters
\be
\la n\ra=\frac {\Gamma}{e^{x}-1}=\beta^2
\ee
and 
\be
e^{-\lambda}=\frac{\beta^2}{1+\beta^2}\;,
\ee
so that the probability distribution of Hawking radiation can be written as
\be
p(n)=\frac1{1-\beta^2}\left(\frac{\beta^2}{1+\beta^2}\right)^n\;.
\ee
Introducing further 
\be 
z=\frac{\beta^2}{1+\beta^2}
\ee
allows us to write
\be 
p(n)=(1-z)z^n\;, \label{phawking}
\ee
which makes the correct normalization of the distribution apparent owing to the geometric series expansion.
\subsection{Stimulated emission}
Bekenstein and Meisels showed that this distribution $p(n)$ cannot consistently describe a black hole that is in equilibrium with its surrounding radiation. Specifically, they showed that if a black hole was immersed in a heat bath with temperature $T$ with distribution $p_\star(n)=(1-e^{-y})e^{-ny}$ where $y=\omega/T$, then the outgoing radiation
\be
p_o(n)=\sum_{m=0}^\infty p(n|m)p_\star(m)\; \label{outdist}
\ee
does not satisfy the detailed balance condition
\be
e^{-xm}p(n|m)=e^{-xn}p(m|n) \;.  \label{detbal}
\ee
if $p(n|m)$, the probability that $n$ particles are outgoing if $m$ particles are impinging on the black hole, is due {\em only to scattering} with a probability $1-\Gamma_0$. We can see this by writing out the probability distribution $p(n|m)$ due to scattering: the binomial
\be
p(n|m)={{m}\choose{n}}\Gamma_0^{m-n}(1-\Gamma_0)^n\;. \label{scat}
 \ee
It is easily checked that Eq.~(\ref{scat}) does not respect the detailed balance condition. As a consequence, such a black hole could not be in equilibrium with its surrounding radiation.

\begin{figure}[htbp] 
   \centering
   \includegraphics[width=2in]{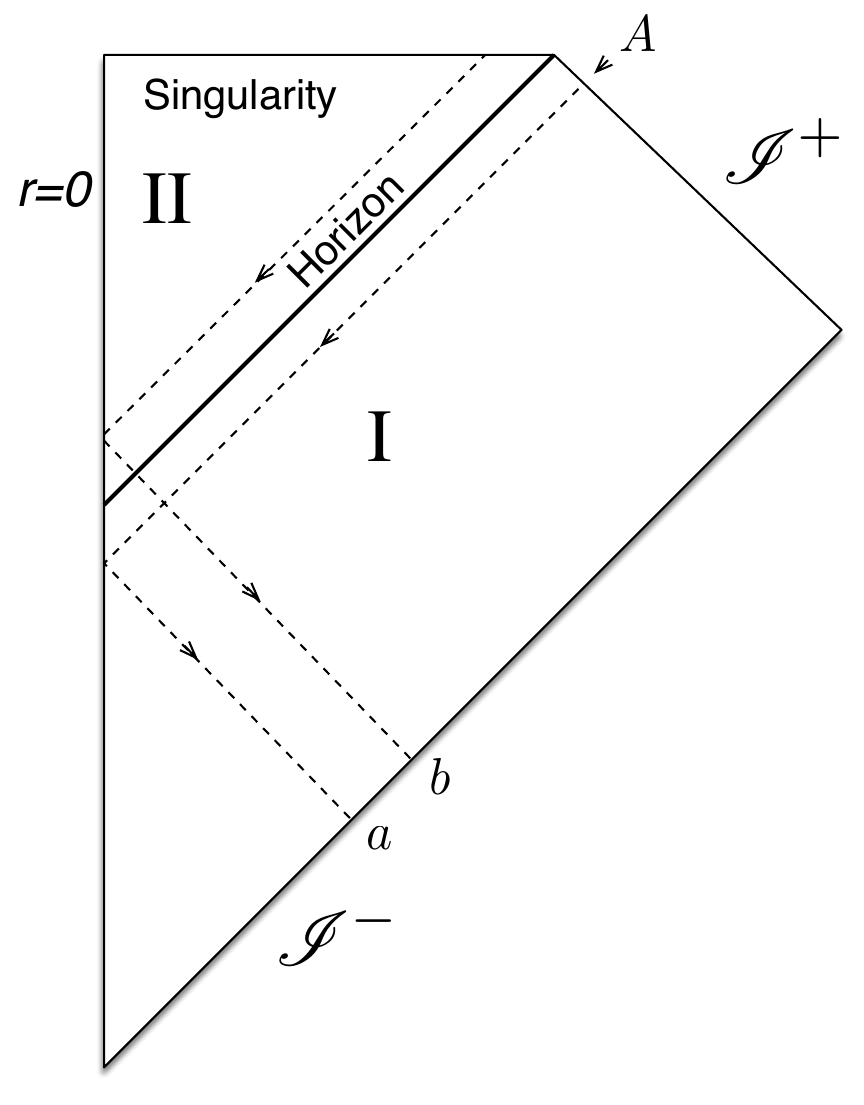} 
   \caption{Penrose diagram showing the early-time modes $a$ and $b$, propagated backwards in time from future infinity ${\mathscr I}^+$ towards past infinity ${\mathscr I}^-$ (just as the black hole formed) and traveling just outside and just inside of the event horizon. The mode at future infinity is annihilated by $A$. In this depiction, modes $a$ and $b$ are shown separated for clarity, but they are just inside and just outside the horizon and coincide at ${\mathscr I}^-$.}
   \label{fig2}
\end{figure}

Stimulated emission can be studied by introducing incoming particles at the exact moment of the formation of the black hole (see Fig.~\ref{fig2}).
%Fig1
Note that in this diagram, the modes $a$ (particles propagating just outside the horizon) and $b$ (propagating just inside the horizon) are followed backwards towards ${\mathscr I}^-$ from ${\mathscr I}^+$, even though we would like to describe the time-reversed situation, where a mode $a$ is scattered towards ${\mathscr I}^+$. In his description, Hawking eliminated positive frequency modes at ${\mathscr I}^-$ after following them backwards from ${\mathscr I}^+$, preventing him from studying stimulated emission. Here, we will study their fate in forward time. This is achieved by the unitary transformation (here and before $\hslash$ is set to 1)
\be
|\Psi\ra_{\rm out}=e^{-iH}|m\ra_a|0\ra_b\;, \label{trans}
\ee
where $H$ implements the Bogoliubov transformation relating in-modes to out-modes (for simplicity we treat only a single mode here, see~\cite{AdamiVersteeg2014,Adami2024} for expressions including a sum over modes)
\be
H=ig(a^\dagger b^\dg-ab) + {\rm h.c.}\;. \label{ham}
\ee
In (\ref{ham}), $g$ is a coupling strength that is related to the black hole surface gravity and sets the temperature of the black hole, and we added the hermitian conjugate in order to be able to treat anti-particles as well. This is not strictly necessary, but is convenient to encode logical bits, as we will see later\footnote{As a consequence, Hamiltonian (\ref{ham}) has an eigenspace that is a product of particle- and anti-particle space, so the in-vacuum in (\ref{trans}) should strictly be written $|m,0\ra_a|0,0\ra_b$.}.

Hamiltonian (\ref{ham}) has the same form as the so-called ``squeezing Hamiltonian" in quantum optics that describes optical parametric amplification~\cite{Leonhardt2010}. Indeed, the black hole is a quantum amplifier as well.

We can calculate the density matrix of outgoing radiation in region I (outside of the black hole horizon, see Fig.~\ref{fig2}) using the formalism of curved-space quantum field theory in the case of full absorption ($\Gamma=1$) using (\ref{trans}) to calculate $\rho_{\rm out}=|\Psi\ra_{\rm out}\la\Psi|$  and tracing over the horizon modes~\cite{AdamiVersteeg2014}.
The resulting outgoing density matrix $\rho(m)$  factorizes into particle and anti-particle matrices
\be \rho(m)&=& \rho_m\otimes \rho_0\;. \label{dens1}
\ee
The matrix $\rho_0$ is just the standard density matrix of Hawking radiation since no anti-particles are sent into the black hole 
\be
\rho_0=\sum_{n=0}^{\infty} p(n)|n\ra\la n|
\ee
with $p(n)$ given by Eq.~(\ref{phawking}), while~\cite{AdamiVersteeg2014}
%\be
%\rho_m=\frac1{(1+\beta^2)^{m+1}}
%\sum_{n=0}^\infty\left(\frac{\beta^2}{1+\beta^2}\right)^{n}{{m+n}\choose{n}}|m+n\ra\la m+n|\;,
%\ee
\be
\rho_m=(1-z)^{1+m}\sum_{n=0}^\infty z^n {n+m \choose n}|n+m\ra \la n+m|
\ee
with $z=\frac{\beta^2}{1+\beta^2}$ as before.

The matrix $\rho_m$ can be rewritten as 
\be
\rho_m=\sum_n p(n|m) |n\ra\la n|\;,
\ee 
with
\be
p(n|m)=(1-z)^{1+m}z^n{n+m \choose n}\;.\label{prob}
\ee
This is the generalization of the spontaneous emission formula to allow for $m$ incoming particles, and reduces to $p(n)$ for $m=0$. We can use those distributions to calculate the mean number of particles outside the horizon. The number of particles due to spontaneous emisison (Hawking radiation) is
\be
\la n\ra_{\rm spont}=\sum_{n=0}^\infty np(n)=\beta^2\;,
\ee
which is the celebrated Planck distribution as
\be
\beta^2=\frac{e^{-\omega_k/T_\bh}}{1-e^{-\omega_k/t_\bh}}\;.
\ee  
The total number of particles emitted in response to $m$ incoming particles is
\be
\la n\ra_{\rm tot}=\sum_{n=0}^\infty np(n|m)=\beta^2(m+1)\;. \label{stim}
\ee
Equation (\ref{stim}) confirms that beyond the Hawking radiation (``+1"), the region outside the horizon has $m\beta^2$ particles that were stimulated by the $m$ incoming particles, with the same spectrum as Hawking radiation. However, while the spectrum is the same, the stimulated particles carry the imprint of the information that disappeared behind the horizon. This should not be surprising to those who realize that stimulated emission is essentially a quantum cloning process~\cite{LamasLinaresetal2002}.

\begin{figure}[htbp] 
\centering
   \includegraphics[width=\textwidth]{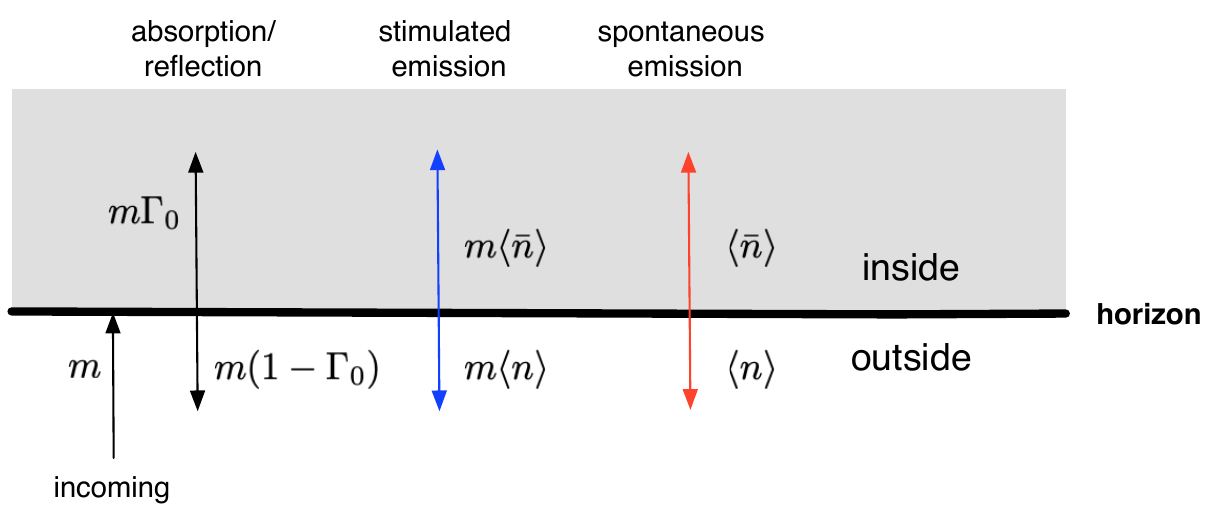} 
   \caption{Outgoing particles (arrows pointing down) for $m$ incident particles on the horizon. In this sketch, $\la n\ra$ ($\la \bar n\ra$) refer to the mean number of particles (anti-particles) given by Eq.~(\ref{hawking}) that are generated in front and behind the horizon (to conserve particle number).}
   \label{fig1}
\end{figure}

Fig.~\ref{fig1} shows a sketch of the particle numbers in front and behind the horizon, as calculated using the unitary transformation (\ref{trans}). For that figure, we have reinstated the ``grey-body" factor $\Gamma_0$, which is the probability for a single quantum to be absorbed at the horizon. This modifies the number of particles in region I to
\be
\la n\ra_{\rm tot}=(1-\Gamma_0)m +\beta^2(m+1) \label{out}
\ee
with 
\be
\beta^2=\frac{\Gamma}{e^{x}-1}=\frac{\Gamma_0(1-e^{-x})}{e^{x}-1}\;.
\ee
The term $(1-\Gamma_0)m$ refers to the number of particles reflected at the black hole horizon. Reflection at the horizon is natural for particles with non-zero angular momentum with respect to the black hole. Only $s$-waves do not scatter.
%Fig2
We will now see whether the stimulated particles outside the horizon are sufficient to reconstruct any classical information that was attached to the particles absorbed at the horizon.

\section{The Holevo Capacity of Black Holes}
The Holevo capacity of a channel~\cite{Holevo1998} is the capacity to transmit classical information over a quantum channel. A positive capacity implies that classical information can be sent over this channel with {\em arbitrary} accuracy, that is, with zero information loss. Just as in the classical theory~\cite{Shannon1948}, the capacity represents the rate of perfect information transfer when information enters the channel at one bit per second. Thus, a channel with a capacity of 0.1 $s^{-1}$, for example, makes it possible to transmit information with perfect accuracy at a rate of 0.1$s^{-1}$. The loss in speed is always due to the error-correcting embedding that is necessary in order to recover the information with perfect accuracy. Let us calculate the Holevo capacity for black holes~\cite{AdamiVersteeg2014}.
\subsection{Early-time Capacity}
We will first calculate the capacity of the black hole channel when information is entering the black hole just after its formation. This is the situation usually represented by the Penrose diagram shown in Fig.~\ref{fig2}, but it is admittedly an awkward channel: nobody would want to transmit information by timing a packet of particles so that it is absorbed just as the black hole is formed. The following calculation is, in fact, just an instructive exercise that will help us understand the more general calculation that comes later, when we send information into an already-formed black hole. 

We imagine that a sender $X$ prepares classical bits using a dual-rail encoding, sending $m$ particles into the black hole signifying a logical `1', and instead $m$ anti-particles when signaling a `0'.   
I displayed the total density matrix outside the horizon earlier as Eq.~(\ref{dens1}): it has simulated emission (which comprises spontaneous radiation) in the particle sector, and only spontaneous emission in the anti-particle sector\footnote{There are other ways to encode information in this channel, even if there were no anti-particles: you could send $m$ particles to signify a `1', and zero particles (``silence") to signify a `0'. This will result in a much lower capacity, however, as error correction is significantly more difficult.}

The distribution $p(n|m)$ shown in Eq.~(\ref{prob}) suffices to calculate the Holevo capacity $\chi$ of the channel. Technically, the Holevo capacity is the maximum Shannon information shared between preparer $X$ (who sends `1'  with probability $p$ and `0' with probability $1-p$) and the area outside the horizon, labeled I in Fig.~\ref{fig2}: $\chi = \max H(X;{\rm I})$. It is thus the generalization of the Shannon capacity, but for quantum channels. The information that $X$ has about region I then takes the simple form~\cite{AdamiVersteeg2014}
\be
\chi=\max_p \chi(p)=\max_p \biggl[S(\rho)-\Bigl(pS[\rho(m)]+(1-p)S[\rho(\bar m)]\Bigr)\biggr]\;. \label{holevo}
\ee
In this expression, $\rho$ refers to the density matrix of the sender $X$: a probabilistic mixture of the 
density matrix for a logical `1' $\rho(m)$, and for the logical `0', $\rho(\bar m)$ using the mixing probability $p$ 
\be
\rho=p\rho(m)+(1-p)\rho(\bar m)\;.
\ee 
Further, $S(\rho)$ is the von Neumann entropy
\be
S(\rho)=-\Tr\,\rho\log \rho\;.
\ee
The von Neumann entropy reverts to the Shannon entropy for diagonal (that is, classical) density matrices.

Let us proceed to calculate $\chi$, but restrict ourselves to sending a single particle or anti-particle to keep the calculation tractable. Setting $m=1$, we need to calculate the entropies $S(\rho)$, $S(\rho(0))$, and $S(\rho(1))$. Now, because $S[\rho(m)]=S[\rho(\bar m)]$ it turns out that
\be
\chi(p)=S(\rho)-S[\rho(1)]\;.
\ee
To calculate $S[\rho(1)]$ we need 
\be
S(\rho_0)=-\sum_{m=0}^\infty p(m|0)\log p(m|0)=-\log(1-z)-\frac{z}{1-z} \log z\;, \label{rho0}
\ee
as well as
\be 
S(\rho_1)=-\sum_{m=0}^\infty p(m|1)\log p(m|1)=-2\log(1-z)-\frac{2z}{1-z} \log z -(1-z)^2 \Delta\;, \ \ \ \ \ \ \  \label{rho1}
\ee
where
\be
\Delta = \sum_{m=0}^\infty z^m (m+1)\log(m+1)\;. \label{deltasum}
\ee
Equation (\ref{deltasum}) has a closed-form expression in terms of derivatives of the polylogarithm function $Li(s,z)=\sum_{n=0}^{\infty}z^n/n^s$
\be
\Delta=-\frac1z\frac{\del}{\del s}{\rm Li}(-1,z)=-\frac{\del}{\del z}\frac{\del}{\del s}{\rm Li}(0,z)\;.
\ee
using the notation $\frac{\del}{\del s}{\rm Li}(0,z)=\frac{\del}{\del s}{\rm Li}(s,z)\vert_{s=0}$.
Gathering (\ref{rho0}) and (\ref{rho1}) we obtain
\be 
S[\rho(1)]=S(\rho_0)+S(\rho_1)=-3\log(1-z)-\frac{3z}{1-z} \log z -(1-z)^2 \Delta\;.
\ee
Let's further calculate $S(\rho)=S[p\rho(1)+(1-p)\rho(\bar 1)]\equiv S(p)$. We find
\be S(p)&=&
-(1-z)^3\sum_{mm'=0}^{\infty}z^{m+m'}\biggl\{[p(m+1)+(1-p)(m'+1)] \biggr.\nonumber \\
&\times&\log\biggl.\left((1-z)^3z^{m+m'}[p(m+1)+(1-p)(m'+1)]\right)\biggr\} \label{chip}
\ee
It is immediately clear that $S(p)=S(1-p)$ as the replacement $p\leftrightarrow(1-p)$ just relabels $m$ and $m'$. Then 
\be
\max_p \chi(p)=\chi(1/2)\;.
\ee
But let's keep $p$ arbitrary for the moment. We can calculate
\be
S(p)=-3\log(1-z)-\frac{3z}{1-z} \log z+(1-z)^3\Lambda(p)\;,
\ee
where
\be
\Lambda(p)= -\sum_{mm'}(1+m'+p(m-m'))\log(1+m'+p(m-m'))z^{m+m'}\;.
\ee
Thus, the term $-3\log(1-z)-\frac{3z}{1-z} \log z$ cancels from $\chi(p)$, and we have
\be
\chi(p)=S(\rho)-S[\rho(1)]=(1-z)^3\Lambda(p)+(1-z)^2\Delta\;.
\ee
We also note that $\chi(0)=\chi(1)=0$.

\begin{figure}[htbp] 
   \centering
   \includegraphics[width=3in]{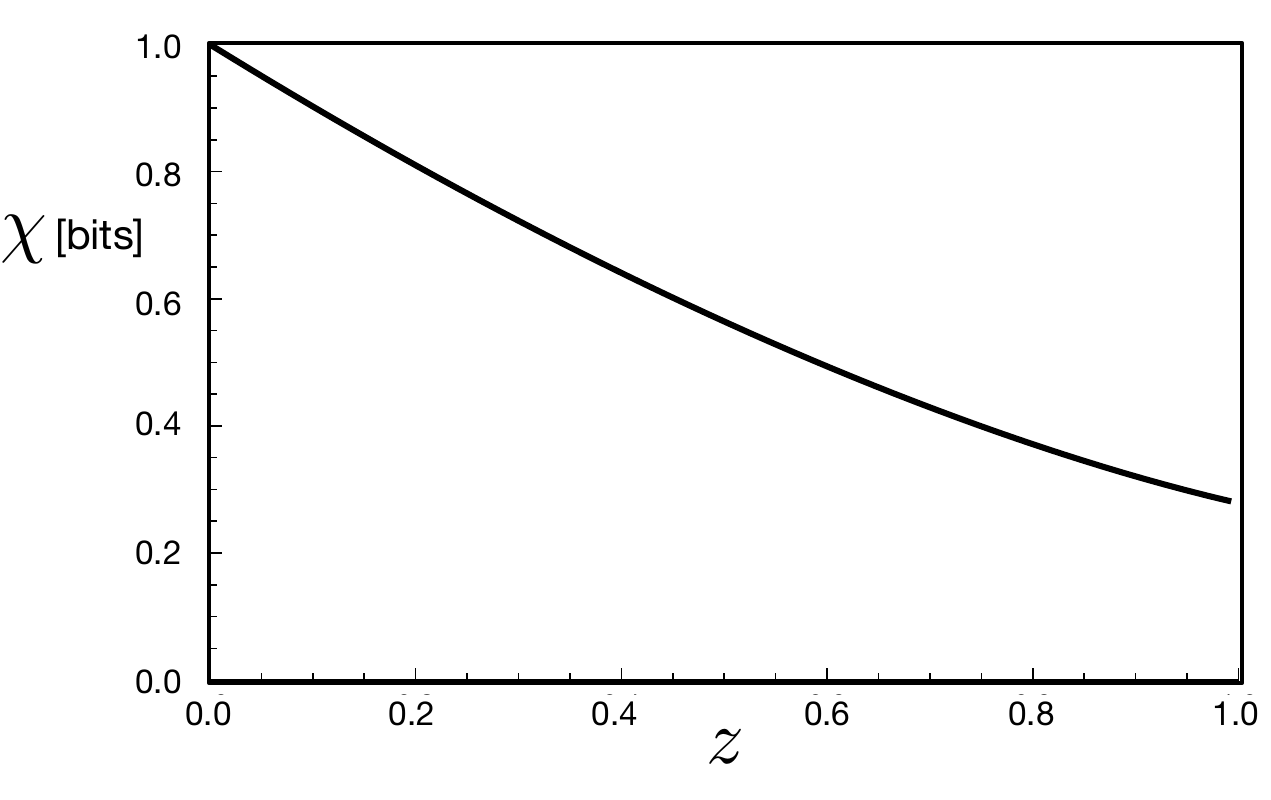} 
   \caption{Early-time capacity $\chi$ for a binary non-reflecting ($\Gamma=1$) black hole channel as a function of the parameter  $z=e^{-\omega/T_{\bh}}$. The limit $z=0$ corresponds to a black hole with infinite mass (vanishing surface gravity), while $z\to1$ as the mass of the black hole tends to zero.}
   \label{cap}
\end{figure}

To obtain the capacity $\chi(1/2)$, we need to calculate $\Lambda(1/2)$. By changing the summation in (\ref{chip}), we can see that
\be
\Lambda(1/2)&=&-\sum_{u=0}^\infty\sum_{v=0}^{u}z^u\frac12(u+2)\log[\frac12(u+2)]\\
&=&-\sum_{u=0}^\infty z^u\frac12(u+1)(u+2)\log[\frac12(u+2)]\\
&=&\frac12\sum_{u=0}^\infty z^u(u+1)(u+2)-\frac12\sum_{u=0}^\infty u(u+1)\log(u+1)z^{u-1}
\ee
where we just shifted the sum in the 2nd term.
Thus,
\be
\Lambda(1/2)=\frac{1}{(1-z)^3}-\frac12\frac1z\sum_{u=0}^\infty z^u u(u+1)\log(u+1)\;.
\ee and therefore (renaming $u\to m$)
\be\chi(1/2)=1-\frac12(1-z)^3\frac1z\sum_{m=0}^\infty z^m m(m+1)\log(m+1)+(1-z)^2\Delta
\ee
Now, 
\be\frac1z\sum_{m=0}^\infty z^m m(m+1)\log(m+1)=-\frac{\del^2}{\del z^2}\frac{\del}{\del s}{\rm Li}(0,z)
\ee
so
\be
\chi(1/2)&=&1+\frac12(1-z)^3\frac{\del^2}{\del z^2}\frac{\del}{\del s}{\rm Li}(0,z)-(1-z)^2\frac{\del}{\del z}\frac{\del}{\del s}{\rm Li}(0,z)\\
&=&1-\frac12(1-z)^2 \frac{\del^2}{\del z^2}(z-1)\frac{\del}{\del s}{\rm Li}(0,z)\;. \label{capearly}
\ee
This is the ``early-time capacity" of the black hole, and its functional dependence on $z$ is shown in Fig.~(\ref{cap}). 
We note that the capacity is everywhere positive, which implies that no information is lost within the black hole ever: it is possible to retrieve any information sent into the black hole from the radiation outside the horizon with arbitrary accuracy.

\subsection{Late-time Capacity}
While a calculation of the early-time capacity is instructive, we also need to acknowledge that it is inconsistent. Indeed, the conditional probability distribution $p(n|m)$ from Eq.~(\ref{prob}) that went into the calculation of the early-time capacity does not obey the detailed balance condition (\ref{detbal}). This is easily seen by noting that $z$ in (\ref{prob}) is just $z=e^{-x}$ (because we had set $\Gamma=1$), and then\footnote{We don't include the binomial coefficient ${n+m \choose n}$  here because it is symmetric in $n$ and $m$.}
\be
e^{-mx}(1-e^{-x})^{m+1}e^{-nx}\neq e^{-nx}(1-e^{-x})^{n+1}e^{-mx}
\ee
The reason for this failure is simple: the use of a black hole's absorptivity $\Gamma=1$ is unphysical as this would imply that no radiation is emitted back via stimulated emission. Indeed, Bekenstein and Meisels found that in order for detailed balance to be respected, the effective absorptivity of the black hole must be
\be
\Gamma =\Gamma_0(1-e^{-x})\;, \label{eff}
\ee
where $\Gamma_0$ is the ``naked" absorptivity of the black hole, that is, the absorptivity determined solely from scattering (see Eq.~(\ref{scat}) and Fig.~\ref{fig1}), so that $\Gamma$ is strictly smaller than 1. 

\begin{figure}[htbp] %  
   \centering
   \includegraphics[width=2in]{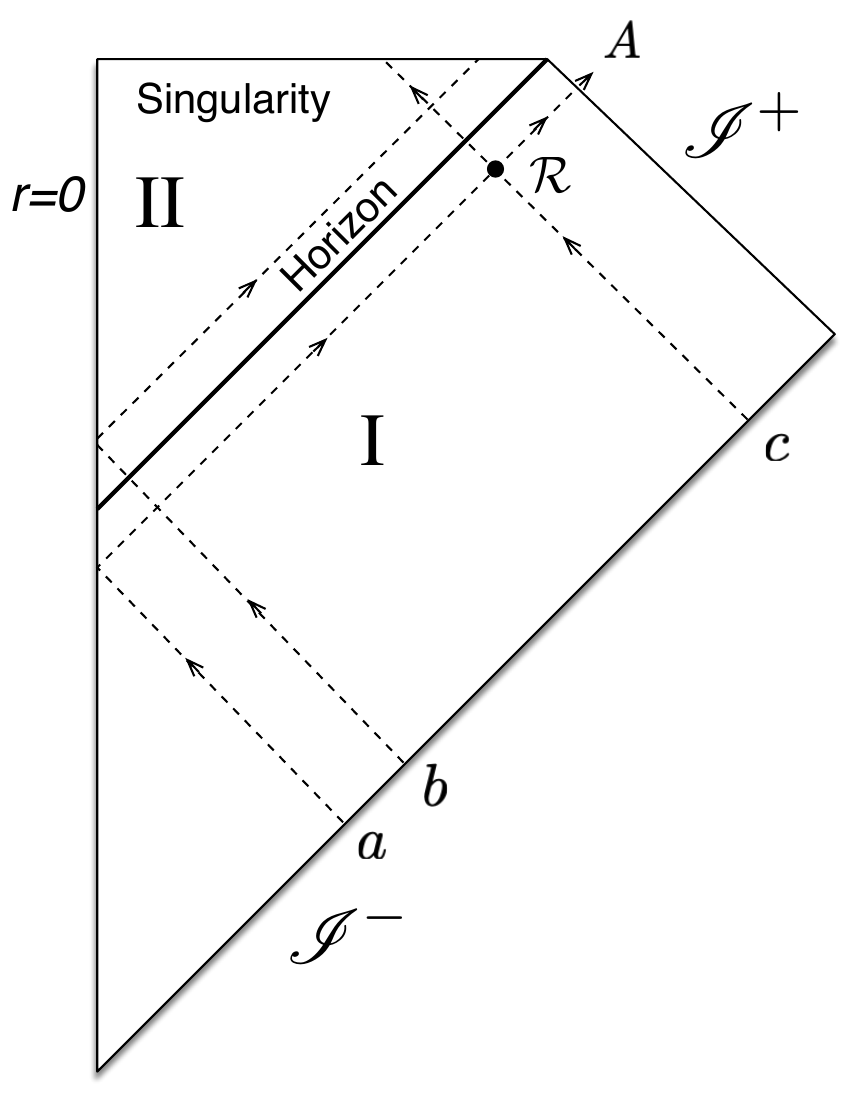} 
   \caption{Penrose diagram showing early-time modes ($a$ and $b$) and late-time modes $c$. Late-time modes are scattered at the horizon with probability ${\pazocal R}=1-\alpha=1-\Gamma_0$ (the black hole reflectivity). A perfectly absorbing black hole has ${\pazocal R}=0$.}
   \label{penlate}
\end{figure}

The correct way to describe scattering with probability $1-\Gamma_0$ is to introduce particles sent into the black hole long after formation of the black hole, instead of creating the ludicrous construction where particles in mode $a$ are sent into the black hole precisely at the moment where it had just formed (Fig.~\ref{fig2}). Sorkin~\cite{Sorkin1987} showed that late-time particles can be consistently described by introducing particles in mode $c$ whose creation and annihilation operators commute with the early-time modes $a$ and $b$ because the $c$-modes are blue-shifted with respect to the early-time modes\footnote{Note that we have changed notatiom from that of Sorkin's (switching $a$ and $c$) in order to be consistent with previous notation.}. The early-time Bogoliubov transformation that links the outgoing annihilation operator $A$ at future infinity (see Figure~\ref{fig2}) to the operators $a$ and $b$ at past infinity is
\be
A=\alpha a-\beta b^\dg \label{bogol}
\ee
with the normalization condition $\alpha^2-\beta^2=1$ (and implemented by the Hamiltonian (\ref{ham}) via $A=e^{-iH}ae^{iH}$)
is then extended to
\be 
A=\alpha a-\beta b^\dg +\gamma c\label{bogol2}\;,
\ee
corresponding to the modes shown in the Penrose diagram in Fig.~(\ref{penlate}).
%Fig3
As the three modes $a$, $b$, and $c$ are orthogonal, the late-time normalization condition is $\alpha^2-\beta^2+\gamma^2=1$. We introduced this notation earlier (see Eq.~(\ref{norm})) without explanation why such a parameterization allowed us to succinctly rewrite the probability distribution of Hawking quanta in a general fashion, for arbitrary absorption coefficient. We now understand the reason why: it is the only way to consistently describe scattering at the black hole horizon. Introducing a scattering rate $1-\Gamma$ when following particles backwards from future infinity towards past infinity (as Hawking did) is inconsistent in more ways than one: when reversing time, we also need to flip the absorptivity of the black hole (more on time reversal later). 

In this parameterization, the coefficient $\alpha^2$ does not refer to the effective absorptivity of the black hole, but rather to the absorption coefficient due to scattering only, that is, $\alpha^2=\Gamma_0$. As before, $\beta^2$ controls the rate of spontaneous emission, but the effective absorption coefficient is $\Gamma=1-\gamma^2$, so that the normalization condition $\alpha^2-\beta^2+\gamma^2=1$ is exactly equivalent to Bekenstein and Meisel's relation (\ref{eff}), which, in turn, is just the familiar relationship between the effective absorptivity of matter $\kappa$ and the opacity $\kappa_0$ for a medium in thermodynamic equilibrium according to the theory of radiative transport (see, e.g.,~\cite{CoxGiuli1968}). 

A perfectly absorbing black hole then has $\alpha^2=1$ (not $\gamma^2=0$). In that case, $\beta^2=\gamma^2$, which implies that the strength of spontaneous emission of late-time particles is the same as that of early-time particles, modulo the difference in red shift, which we have ignored here. In fact, this allows us to see that the late-time formalism resolves the transplanckian paradox (see adjacent box).  

\begin{svgraybox}

\centerline{\bf Resolving the Transplanckian Paradox}
\vskip 0.25cm
We recall that 'tHooft~\cite{thooft1985} and later Jacobson~\cite{Jacobsen1991} pointed out that when the initial modes $a$ and $b$ are transported along the horizon to late times, they acquire an infinite redshift (see also the discussion in~\cite{FabbriNavarro-Salas2005}). Besides leading to conceptual problems associated with a breakdown of the semi-classical approximation, the infinite redshift of early-time modes also contradicts the notion that radiation emanating from an object with the diameter $R=2M_\bh$ (where $M_\bh$ is the black hole mass) should instead have a frequency of $\omega~\sim\frac1{M_{\bh}}$, an essentially infinite discrepancy for large black holes. A popular ``resolution" of this paradox was the so-called ``black hole complementarity" principle~\cite{thooft1990,susskindetal1993}. Black hole complementarity posited that the horizon in reality consists of an ``exterior" and an ``interior" horizon that are complementarity in the sense that states on the surfaces of these two horizons are othogonal (the Hilbert space factorizes into a product of interior and exterior spaces). We can now see that such a construction (which incidentally renders the energy-momentum tensor infinite at the horizon, that is, it is not renormalizable)
is completely unnecessary as it is instead the Hilbert space of early- and late-time modes that factorizes. Because late-time particles do not accumulate a redshift, both the spontaneous and stimulated radiation associated with them is not infinitely redshifted, and their frequency instead is commensurate with the size of the black hole.
\end{svgraybox}

\begin{figure}[htbp] 
   \centering
   \includegraphics[width=3in]{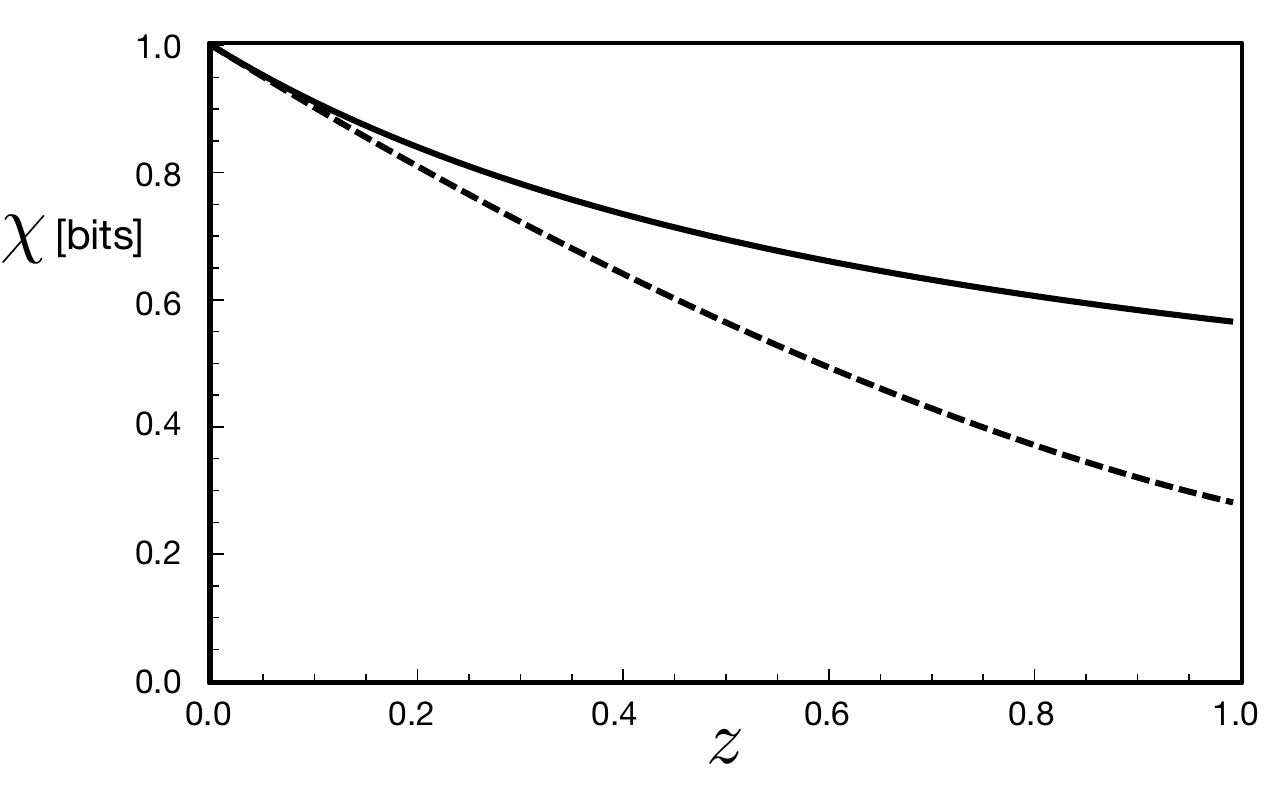} 
   \caption{Capacity $\chi$ for a binary non-reflecting black ($\Gamma_0=1$) hole channel as a function of the parameter  $z=e^{-\omega/T_{bh}}$.  The solid line represents the late-time capacity, while the dashed line is the capacity for early-time modes. Note that since $g\approx z$, we can see this plot also as depicting the dependence of the information transmission capacity on the mode coupling strength in the black-hole Hamiltonian.}
   \label{latecap}
\end{figure}

The direct calculation of $p(n|m)$ in curved-space quantum field theory is complicated and we will not repeat it here (it was performed independently using three different methods in~\cite{PanangadenWald1977,AudretschMueller1992,AdamiVersteeg2014}). The most intuitive calculation writes the unitary transformation from past-infinity vacuum to future-infinity vacuum
\be
|0\ra_{\rm out}=e^{-iH}|0\ra_{\rm in} \label{ham1}
\ee
in terms of a Hamiltonian $H$ that can be written as a sum of two terms: one that is equivalent to an active optical element, the so-called ``squeezing Hamiltonian" (OPA stands for ``optical parametric amplification") with coupling strength $g$
\be
H_{OPA}=ig(a^\dg b^\dg-ab) +{\rm h.c.} \label{squeeze}
\ee
and the Hamiltonian of a passive optical element, the ``beam splitter"
\be
H_S=ig'(a^\dg c-ac^\dg) +{\rm h.c.}\label{scatter}
\ee
that scatters late incoming modes $c$ into horizon modes $a$. In (\ref{scatter}), the scattering strength $g'$ is bounded from below by the coupling strength $g$ and sets the black hole reflectivity via~\cite{AdamiVersteeg2014}
\be
\alpha^2=\cos^2(\sqrt{g'^2-g^2})\;.
\ee
The resulting late-time probability distribution $p(n|m)$ for arbitrary $m$ in terms of $\alpha$, $\beta$, and $\gamma$ can be written as
\be
p(n|m)=(1-z)^{m+1}z^n(\alpha^2)^m\sum_{k=0}^{\min(n,m)} (-1)^k{m\choose k}{m+n-k \choose n-k}(1-\frac{\gamma^2}{\alpha^2\beta^2})^k\;. \ \ \ \ \ \ \   \label{full}
\ee
We can quickly check a few limits. For a perfectly black (absorbing) hole, $\alpha^2=1$ which makes the term in brackets $1-\frac{\gamma^2}{\alpha^2\beta^2}=0$. As a consequence, only the term in the sum with $k=0$ survives, leading to
\be
p(n|m)=(1-z)^{m+1}z^n {m+n\choose n}\;. \label{probdet}
\ee
This distribution looks exactly like the early-time distribution (\ref{prob}), but it is not: here, $z=\frac{\beta^2}{1+\beta^2}$ and it is {\em not}\footnote{In Eq.~(\ref{prob}), $\gamma^2=0$ and therefore $\beta^2+1=\alpha^2$.} equal to $e^{-x}$. Rather, we have
\be
e^{-x}=\frac{z}{1-z}\;,
\ee
which we can use to show that (\ref{probdet}) indeed satisfies detailed balance. Using (\ref{probdet}) to calculate the capacity of a fully absorbing black hole gives us the solid curve in Fig.~\ref{latecap}. The late-time capacity is strictly larger than the early-time capacity (shown as a dashed line in Fig.~\ref{latecap} for comparison), because the late-time capacity can be obtained from the early-time capacity simply by a replacement $z\to\frac{z}{1+z}$~\cite{AdamiVersteeg2014}.
%Fig4

We can also use Eq.~(\ref{full}) to derive the probability distribution of outgoing radiation for a white hole ($\alpha^2=0$). In that case, it is the {\em last} term in the sum that survives (all other terms vanish on account of $(\alpha^2)^m$ in Eq.~(\ref{full})), giving rise to the white hole distribution
\be
p(n|m)=(1-z)^{m+1}z^{n-m}{n \choose m}\;. \label{white}
\ee
We can use this distribution to calculate the number of particles outside the white hole horizon as
\be
\la n\ra=\sum_{n=0}^\infty n p(n|m)=\frac{m+z}{1-z}=m(1+\beta^2)+\beta^2\;.
\ee
This is the result (\ref{out}) with $\Gamma_0=0$.
 
\section{Time-Reversal Invariance}
\begin{figure}[htbp] %  figure placement: here, top, bottom, or page
   \centering
   \includegraphics[width=\linewidth]{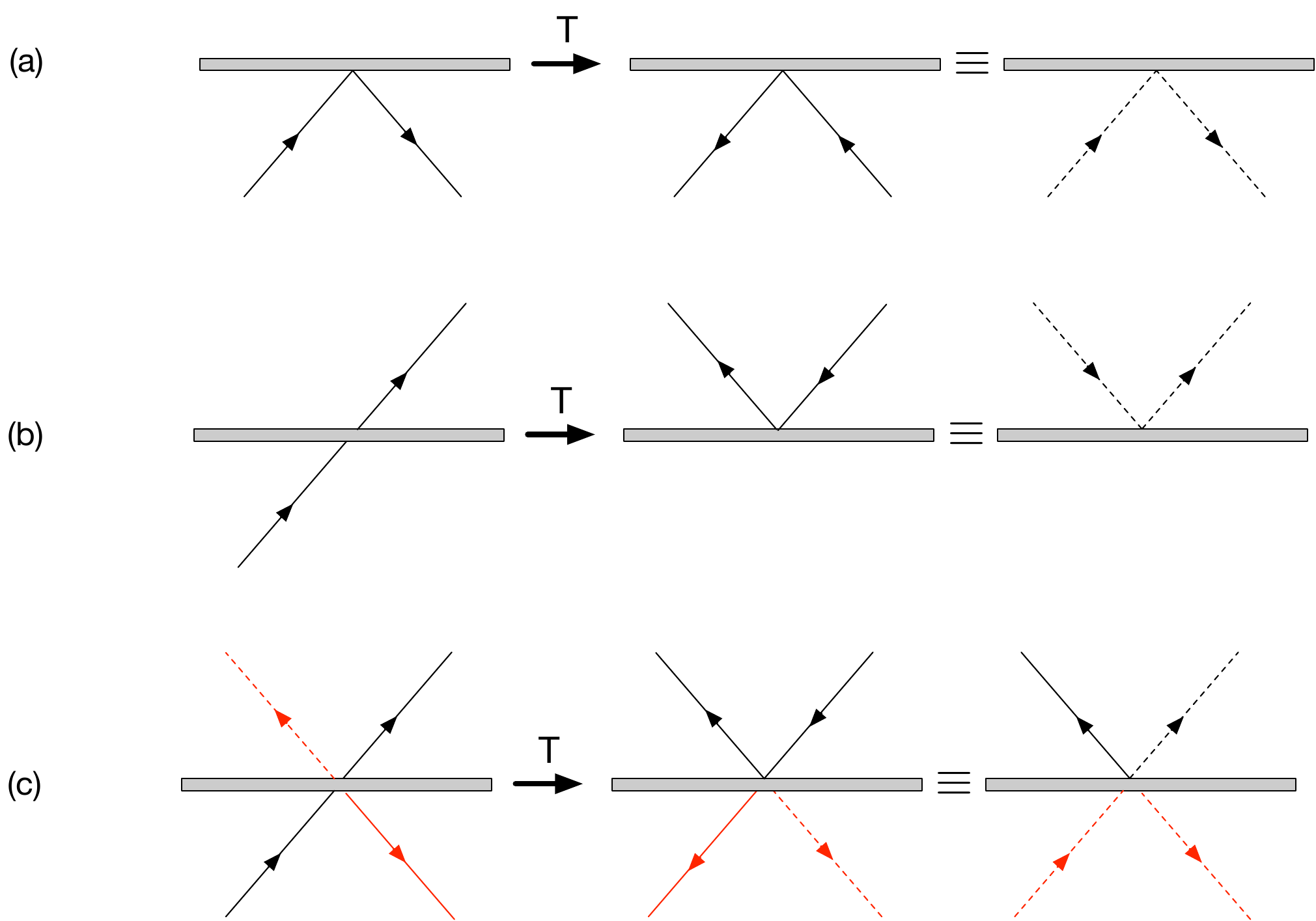}
   \caption{Sketch of particle and anti-particle trajectories in the presence of white- and black hole horizons. Incoming particles are in black, while stimulated particles/anti-particles are in red. Solid lines represent particle trajectories, while dashed lines illustrate anti-particles. (a) Classical trajectory ``reflected" at a mirror (i.e., falling back under the influence of gravity) on the left, its time-reversed trajectory (middle), and the equivalent trajectory in which particles moving forward in time are replaced by anti-particles moving backwards in time (right). (b) Classical trajectory of a particle absorbed by a black hole horizon (left), its time-reversed trajectory (middle), and the CP-transformed process (right). (c): Quantum trajectories including stimulated emission effects. The time-reversed process is equal to the CP-transformed process where particles are replaced by anti-particles moving backwards in time.\label{TR}}
\end{figure}

In the previous discussion, I briefly alluded to the problem of time reversal in the presence of black holes. Indeed, it is possible to see the entire black hole information paradox as one that breaks probability conservation due to a failure of time-reversal invariance~\cite{Witten2012}. Newtonian gravity is, of course, invariant under time-reversal at the micro-level: every particle trajectory in a gravitational field, when time-reversed, gives rise to another plausible particle trajectory. One way to illustrate this is as in Fig.~\ref{TR}a (left panel), where (to simplify the discussion) the effect of a gravitational field ``pulling down" a particle is shown as if a perfect mirror reflected the particle (instead of showing a parabolic trajectory).
%Fig5
When reversing time, the trajectory simply reverses (middle panel). Using charge-conjugation symmetry (replacing a particle moving forward in time by an anti-particle moving backwards in time) gives us the right panel. Comparing the first and the last panel, we see that time reversal is equivalent to CP symmetry, which is the content of the celebrated CPT theorem~\cite{Streater1964}. 

It is easy to see that classical black holes violate this theorem. If instead of falling back a particle instead encounters a perfectly absorbing black hole horizon (Fig.~\ref{TR}b, left panel) reversing the arrow of time (middle panel) does {\em not} generate the time-reversed picture of the left panel, since a particle from inside of the black hole cannot escape to the outside. Instead, it is ``reflected" at the horizon. Indeed, viewed from inside a black hole must act like the perfect mirror in Fig.~\ref{TR}a, as a particle cannot escape past the horizon. In other words, viewed from the inside a black hole must look like a white hole. Replacing particles by anti-particles moving backwards in time produces the right panel, which evidently is {\em not} the anti-particle version of the left panel, thus breaking CPT invariance. 

Simply adding Hawking radiation to this picture does not restore CPT invariance. However, including stimulated emission of radiation does. We can see in Fig.~\ref{TR}c  the stimulated particle/anti-particle pair in red (full absortion, i.e., $\Gamma_0=1$ is shown here) that must accompany the absorption process. Time-reversing this trajectory produces the middle panel, where the particle from inside of the black hole indeed reflects at the horizon, but it {\em also} stimulates a particle/anti-particle pair outside of the horizon (shown in red). In fact, this is the white-hole stimulated emission process described earlier via Eq.~(\ref{white}). Re-interpreting particles moving backwards in time as anti-particles moving forwards in time produces the right panel in Fig.~\ref{TR}c, which indeed is the same process as in the left panel, only with particles replaced with anti-particles. Thus, stimulated emission of pairs restores CPT invariance. I have not shown spontaneous emission of pairs (Hawking radiation) in these diagrams as those particles have no influence on CPT invariance: they only serve to safeguard the no-cloning theorem.

\section{Discussion}
That something was not right with classical black holes was clear early on (as they manifestly violate time-reversal invariance). Witten pointed out~\cite{Witten2012} that not only classical but also quantum black holes violate time-reversal invariance (along with detailed balance), and this is perhaps the most succinct way of characterizing the ``information paradox" discussed in this work. It is now clear that this problem is entirely due to a mistake of Hawking. Made aware of superradiance by Zeldovitch and Starobinskii, he realized (as Zeldovitch and Starobinskii had suggested to him) that black holes must also emit {\em spontaneously}. Unaware that there is a quantum phenomenon of superradiance that does not require negative effective frequencies, Hawking simply put to zero amplitudes at past infinity (propagating modes backward from future infinity) because he was treating Schwarzschild black holes that cannot superradiate, unaware of quantum superradiance (stimulated emission).  

We might ask why Hawking chose to follow modes backwards in time, rather than forward in time as we do here. I believe the answer is that doing so is awkward: if a mode is released before the formation of the black hole, then it could not be ``reflected though the center"~\cite[p. 208]{Hawking1975}. But if it was released after the formation of the hole, it would have to be scattered at the horizon, not by the center. Indeed, our calculation of the early-time capacity using particles released with precise timing towards the soon-to-be-forming horizon mirrors that awkwardness: the resulting radiation does not conform to detailed balance and therefore is not physical. The solution to this conundrum was found by Sorkin~\cite{Sorkin1987}, who introduced a ``late-time" Hilbert space that decouples from the in-vacuum Hilbert space that defines the forming black hole. The resulting equation for the probability distribution of outgoing radiation, using conventional curved-space quantum field theory methods, is fully consistent with that obtained by purely statistical (maximum entropy) arguments by Bekenstein and Meisels~\cite{BekensteinMeisels1977}. It is fair to say that this problem is now solved for all time. 

\begin{figure}[htbp] 
   \centering
   \includegraphics[width=3in]{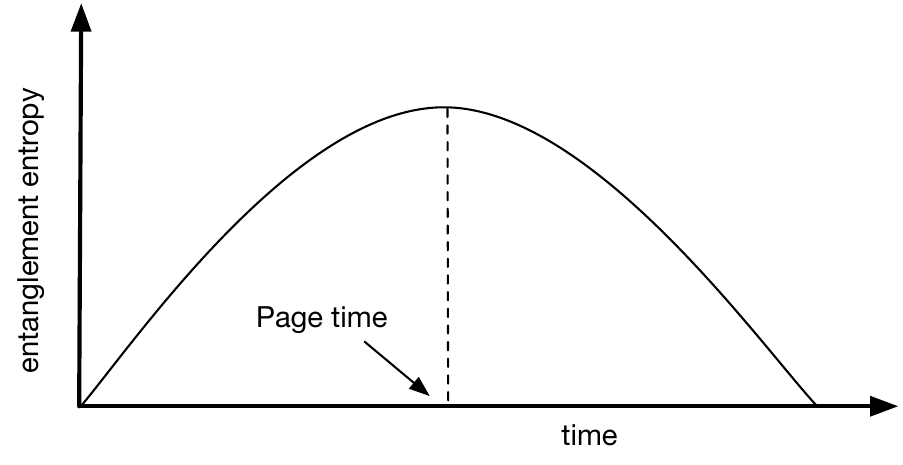} 
   \caption{The Page curve examines the relationship between the entanglement entropy of Hawking radiation and time (strictly speaking, the "size" of the subsystem as determined by the number of particles in the subsystem). In a unitary picture of black hole formation and evaporation, the entanglement entropy should be rising up to the Page time, and then fall again until ultimately reaching zero.}
   \label{page}
\end{figure}

However, all might still not be well with black holes. A second paradox (which is different from the ``information paradox") could still be plaguing black hole physics. This problem concerns the unitarity of the evaporation process of the black hole (as pointed out early on by Hawking~\cite{Hawking1976b}), and the apparent ``lack of predictability" is also related to a failure of time-reversal invariance. However, this problem is not solved by stimulated emission of radiation, as the loss of prediction is not due to losing information that is absorbed by a black hole, but the predictability of space-time itself. The problem with a unitary time-evolution of a black hole from formation to evaporation is best exemplified by the ``Page curve"~\cite{Page1993}, which we can interpret as a curve showing the entropy of the black hole as a function of time (shown in Fig.~\ref{page}).
In a unitary picture of black hole dynamics, a black hole that started out as a quantum mechanical pure state (with vanishing entanglement entropy) should end up as a pure state again, as opposed to a mixed state with positive entropy. It is difficult to calculate the entanglement entropy of a black hole as a function of time because a calculation of the black hole $S$-matrix (which tracks dynamics over time) requires a Hamiltonian that describes the interaction between the black hole degrees of freedom and Hawking radiation. However, curved-space quantum field theory does not yield such an interaction Hamiltonian: the standard theory (even including the scattering term (\ref{scat})) is a free-field theory. 

One way out of this conumdrum is to take the quantum optics analogy seriously, and look towards interaction Hamiltonians whose limit is the free field Hamiltonian (\ref{ham}). Indeed, the squeezing Hamiltonian is the limit of a more general Hamiltonian that takes into account the degrees of freedom of the pump (typically a laser) that powers the creation of entangled pairs (called ``signal" and ``idler" modes). In the limit (\ref{ham}) this pump is usually assumed to be so intense that the average number of photons in this mode is constant: the pump is said to be {\em inexhaustible}, meaning that the source can produce entangled pairs of particles endlessly. Realistic pumps, however, are not ``undepletable": the production of pairs can ``back-react" on the pump leading to its exhaustion.  The quantum Hamiltonian that couples the pairs to the pump degrees of freedom is well-known~\cite{TuckerWalls1969a,TuckerWalls1969b,WallsBarakat1970,AgrawalMehta1974}, and accurately describes realistic experimental situations. Not surpringly, such ``trilinear" Hamiltonians have been used to model the interaction between black hole modes (standing in for the pump modes, since in this picture the black hole is the quantum amplifier) and the horizon modes $a$ and $b$~\cite{NationBlencowe2010,Alsing2015} . The ``backreaction" of the Hawking radiation on the black hole itself is thus modeled precisely in the manner in which the production of signal/idler pairs depletes the pump, until the mean number of photons in the pump is zero (the black hole has disappared).  

Using such a trilinear interaction in a Monte Carlo path integral approach to the $S$-matrix, it is possible to study formation and decay of a black hole with a fixed mass (fixed number of pump modes), giving rise to Page curves~\cite{BradlerAdami2016,AlsingFanto2016}. In that work, the $b$ modes created behind the horizon were assumed not to interact with the black hole modes anymore, while this is in theory possible. An extended analysis that allows internal $b$ modes to create additional pairs shows that lifting this approximation does not destroy the Page curves~\cite{Alsing2025}. 

Granted, the interaction that makes a unitary time evolution of black holes possible does not follow from a fundamental theory of quantum gravity. However, it can be argued that {\em any} consistent unitary theory that obeys energy conservation rules must have a tri-linear interaction term in the low-energy limit (argued, for example, by Strominger~\cite{Strominger1996}). At the same time, it is remarkable that the process that leads to the ultimate decoupling of black hole degrees of freedom from the radiation degrees of freedom turns out to be the analogue of a well-known quantum information protocol, namely the fully-quantum Slepian-Wolf (FQSW) protocol~\cite{Abeyesingheetal2009} (see the discussion in~\cite{Adami2024}). This too is unlikely to be a coincidence.

We thus come to the conclusion that black holes are interesting astrophysical objects that, rather than upending the laws of physics, have taught us much about the most fundamental lawas of physics, such as time-reversal invariance and the no-cloning theorem. And perhaps they have more surprises left in them after all.  

\begin{acknowledgement}
I am indebted to my collaborators in black hole physics: Greg Ver Steeg and Kamil Bradler. 
\end{acknowledgement}

%\bibliographystyle{plain}
%\bibliography{Springer-BH}

%%%%%%%%%%%%%%%%%%%%%%%%%%%%%%%%%%%%%%%%%%%%%%%%%%%%%%%%%

\end{document}